\documentclass{epl}
\title{Giant Keplerate molecule \chem{Fe_{30}} - the first octopole magnet}
\shorttitle{The first octopole magnet}
\author{A. S. Mischenko\inst{1}\thanks{E-mail: \email{smischenko@yahoo.com}}
\and A. S. Chernyshov\inst{1} \and A. K. Zvezdin\inst{2}}
\institute{
  \inst{1} Physics Department, Moscow Lomonosov State University, 119889, Moscow, Russia\\
  \inst{2} General Physics Institute of RAS, 119991, Moscow, Russia
}
\begin{document}
\maketitle
\begin{abstract}
The multipole expansion technique is applied to one of the largest
magnetic molecules, \chem{Fe_{30}}. The molecule's dipole, toroid
and quadrupole magnetic moments are equal to zero (in the absence
of magnetic field) so the multipole expansion starts from the
octopole moment. Probably the \chem{Fe_{30}} molecule is the most
symmetrical magnetic body synthesized so far. The magnetization
process is considered theoretically in different geometries. Some
components of the octopole moment experience a jump while the
magnetization rises linearly up to its saturation value. An
elementary octopole moment consisting of four magnetic dipoles is
proposed as a hint for designing of an experiment for measurement
of octopole magnetic moment components.
\end{abstract}
\section{Introduction}
Multipole expansion of the electrical field generated by a number
of electrical charges has been widely used in theoretical physics.
However, multipole expansion of magnetic fields generated by a
system of permanent currents is less known. The starting point of
such an expansion is a well known formula for the vector potential
of the magnetic field:
\begin{equation}
\label{e.1} \vec{A} = \frac{1}{c} \sum_{i} { \frac{e _i \vec{v
_i}}{| \vec{R}- \vec{r _i}|}},
\end{equation}
where $e_{i}$ and $\vec{v_{i}}$ stand for a charge and velocity of
the $\emph{i}$-th charge. The vector potential $\vec{A}$ can be
expanded in a power series of the $|\vec{r}| / |\vec{R}|$ ratio if
an interesting point is rather distant from the system of charges,
$i.e.$ $|\vec{r}| / |\vec{R}| \ll 1$. The first term of the
expansion in question is $\vec{A'}=[\vec{M}\vec{R}]/{R^3}$ where
$\vec{M}=\frac{1}{2c}\sum_{i} { e _i [\vec{r _i} \vec{v _i}]}
=\frac{1}{2c}\sum_{i}{\frac{e_i}{m_i} \vec{l_i}}$ is the magnetic
dipole moment of the system, $\vec{l_i}$ are orbital moments of
the electrons/ions and $m_i$ are their masses. The spin
$\vec{s_i}$ degrees of freedom can be taken into account by the
substitution $\vec{l_i} \rightarrow \vec{l_i}+g_i\vec{s_i}$. In
this paper we will be concerned with the spin magnetism only. A
distinctive feature of the magnetic field potential's (\ref{e.1})
multipole expansion is that it is subdivided into two families:
magnetic and toroid moments, unlike the expansion of the scalar
electrical potential. The first and easiest representative of the
toroid family is the anapole moment \cite{b.1} - a torus with
electric currents flowing along its meridians. Magnetic multipole
moments are defined by transverse currents (currents flowing along
the torus' parallels for instance). The necessity of the
subdivision is quite obvious since we can decompose the current
density into the following terms according to a well known theorem
of the vector analysis \cite{b.2}:
$\vec{j}(\vec{r})=grad(\eta)+rot(\vec{f})$, where the first term
corresponds to the toroid moments and the second to the "usual"
magnetic multipole moments. Spin systems of finite sizes and
magnetic molecules (\chem{Mn_{12}Ac, V_{15}, Fe_{8}, Fe_{10},
Fe_{30}}, $etc.$) in particular are natural objects for
application of the multipole expansion technique. The magnetic
molecules have attracted great attention from the point of view of
fundamental problems of quantum mechanics in general and the
theory of magnetism in particular \cite{b.3} as well as
nanotechnology and microelectron applications (as model systems
for quantum informatics for example) \cite{b.3, b.4, b.5, b.6,
b.7} in recent years. The molecules are large organic molecular
complexes with the weight of approximately $10^3$ atomic mass
units. The magnetic ions, such as \chem{Fe, Mn, V}, etc. embedded
within the molecule cause their interesting magnetic properties.
There is a strong exchange interaction between magnetic ions (in
the order of $10^6$ Oe) within the molecule.
\section{Magnetic multipole expansion}
The lowest magnetic multipole moments of spin systems (dipole
$Ì_1$, quadrupole $Ì_2$, octopole $Ì_4$ and toroid $Ò_1$) can be
defined as follows \cite{b.8, b.9, b.10}:
\begin{eqnarray}
\label{e.2a} \vec{M _1} = \mu_B\sum_{i}{g_i\vec{s_i}}, \qquad
{(M_2)}_{\alpha\beta} =
\mu_B\sum_{i}{g_i(r_{i\beta}s_{i\alpha}+s_{i\beta}r_{i\alpha})},
\\
\label{e.2b} {(M_3)}_{\alpha\beta\gamma} =
3\mu_B\sum_{i}{g_is_{i\alpha}r_{i\beta}r_{i\gamma}},
\\
\label{e.2c}
 M_{n,m} =
-\mu_B\sum_{i}{g_i[\vec{\nabla_i}({r_i}^n{C_n}^m(\theta_i,
\varphi_i))\times\vec{s_i}]}, \qquad \vec{T_1} =
\mu_B\sum_{i}{g_i[\vec{s_i}\vec{r_i}]},
\end{eqnarray}
where $M_{n,m}$ are magnetic multipole moment's components in
spherical coordinates ($n$ defines the order of a moment: $n = 1$
- dipole, $n = 2$ - quadrupole, $n = 3$ - octopole, $etc.$, $m$
ranges from $-n$ to $n$ by 1). It is a well-known fact that each
multipole moment interacts with the corresponding magnetic field
spatial variation. For the sake of brevity we present here the
formula for the octopole moment only: $W =
-(1/6)(M_3)_{\alpha\beta\gamma}B_{\gamma\beta\alpha}, \quad
B_{\gamma\beta\alpha} = \nabla_{\gamma}\nabla_{\beta}B_{\alpha}$,
where $W$ is the energy of interaction between an octopole and
external magnetic field in Cartesian coordinates,
$\nabla_{\gamma}$ and $\nabla_{\beta}$ are Laplace operators for
the spatial derivatives and $B_{\alpha}$ stands for the three
components of the external magnetic field.

Octopolar electric fields have already been well examined in
connection with non-linear optical properties \cite{b.new}.
Quadrupole moment of the rare earth ions, their dependence on
external magnetic field and connection with magnetic birefringence
and magnetostriction have been investigated in \cite{b.11, b.12,
b.13}. The spin toroid moment $T_1$ has been experimentally found
in bulk magnets (\chem{GaFeO_3, Cr_2O_3}, etc.) recently
\cite{b.14}. Molecular complexes with the spin density that can be
described by means of the toroid moment $T_1$ have been analyzed
theoretically in \cite{b.15} but an experimental realization of
the proposed idea still remains unknown. Quadrupole magnetic
fields generated by antiferromagnetic crystals have been analyzed
in \cite{b.16}. Antiferromagnetic nanoclusters, $e.g.$
\chem{Fe_{10}, Fe_6} can be characterized by the quadrupole moment
$M_2$ and moments of higher orders.
\section{Highly symmetrical Keplerate molecule \chem{Fe_{30}}}
However, no one magnetic object has been characterized by the
octopole moment so far. The first object of this kind is a
magnetic nanocluster \chem{Fe_{30}} with the chemical formula \\
\chem{[Mo_{72}Fe_{30}O_{252} (Mo_{2O7}(H_2O))_2
(Mo_2O_8H_2(H_2O)) (CH_3COO)_{12} (H_2O)_{91}] * 150 H_2O}. \\
This molecule has the largest number of magnetic $Fe^{3+}$ ions
among all the known molecular magnets synthesized so far
\cite{b.4}. The molecule \chem{Fe_{30}} as well as the other
nanoclusters discussed above occupies an intermediate position
among the bulk materials and localized magnetic ions, that is the
reason these objects are called "mesoscopic magnets". Toroid $(T
_1)$, dipole $(M _1)$ and quadrupole $(M _2)$ moments of this
molecule are equal to zero, $i.e.$ the multipole expansion starts
from the octopole moment. In other words, probably the
\chem{Fe_{30}} molecule with a highly symmetrical spin density is
the most symmetrical magnetic body synthesized so far.

\begin{figure}
\twofigures[scale=0.42]{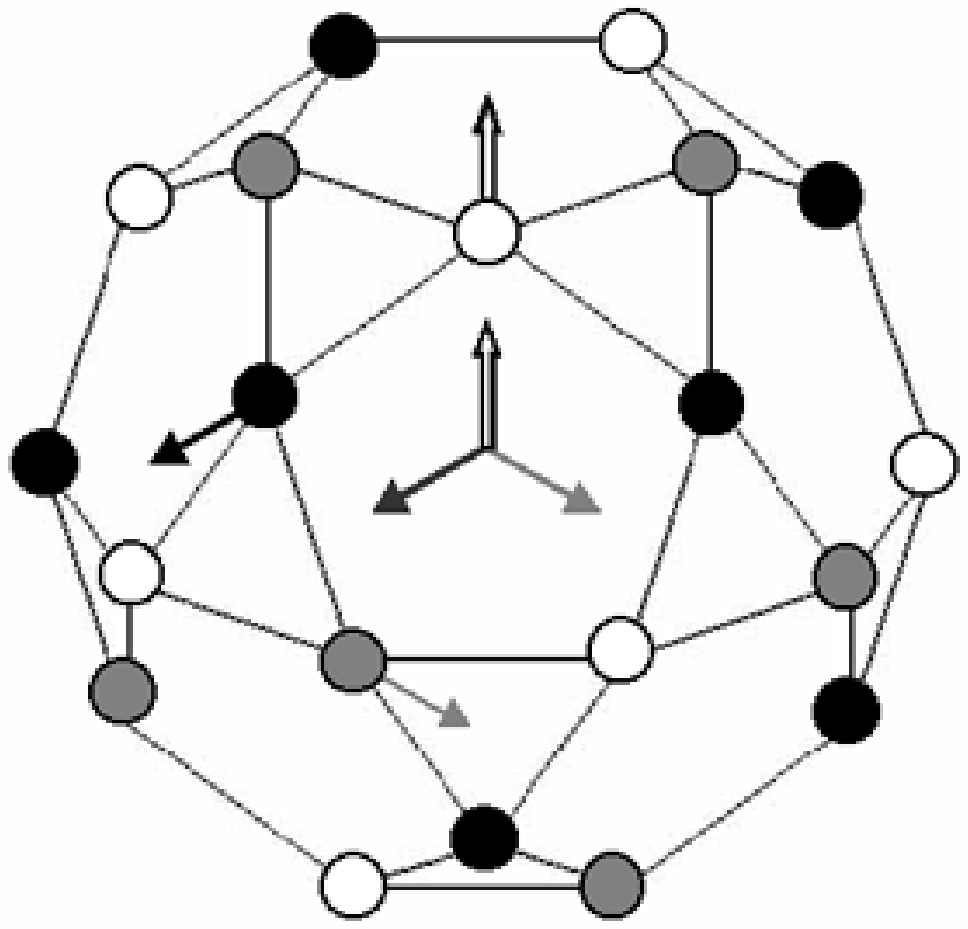}{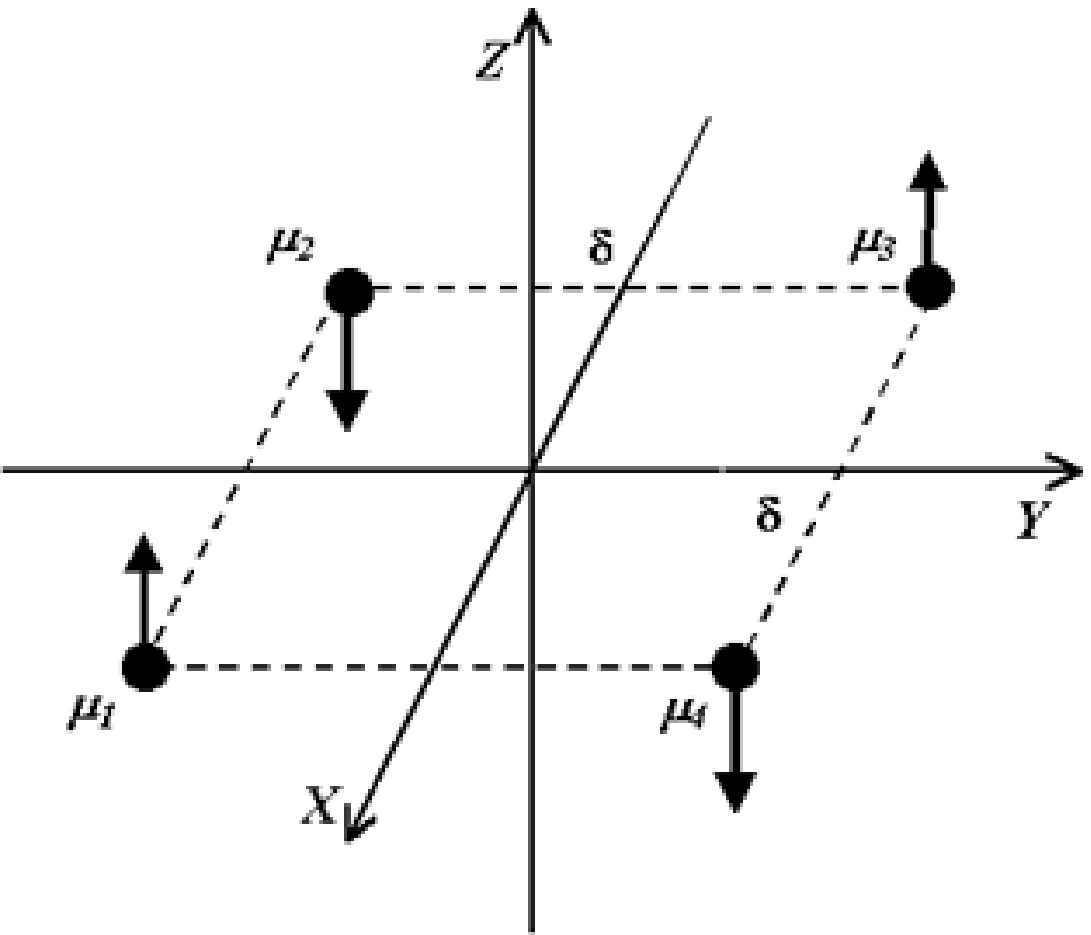} \caption{A schematic
picture of the \chem{Fe_{30}} molecule. The three possible spin
directions are marked by black, gray and white colors.}
\label{f.1} \caption{A model of an elementary octopole consisting
of four equal magnetic moments $m$ situated in the vertices of the
square with a side $\delta$.} \label{f.2}
\end{figure}

The investigated molecular complex consists of 30 $Fe^{3+}$ ions
each of which has a spin $s = 5/2$. The magnetic ions are situated
in the vertices of the icosidodecahedron, one of the regular
Archimedean polytopes (see fig.\ref{f.1}). The icosidodecahedron
has $v = 30$ vertices, $f = 32$ faces and $e = 60$ edges in
accordance with the so-called Euler theorem \cite{b.1}, connecting
a number of vertices, faces and edges of a convex polyhedron of O
genus $(v + f - e = 2)$. The faces of the icosidodecahedron
consist of 20 triangles and 12 pentagons. Neighbouring $Fe^{3+}$
ions interact with each other antiferromagnetically by means of
the indirect exchange interaction which results in zero total spin
of the molecule in the absence of magnetic field \cite{b.7}. The
exact ground state energy and the corresponding ground state of
the \chem{Fe_{30}} molecule has been found in \cite{b.7} basing on
graph-theoretical technique and a special geometrical property of
the icosidodecahedron, its three-colorability (a graph is called
three-colorable if we can color all its vertices with three colors
and no neighboring vertices with the same color - see
fig.\ref{f.1}). In the framework of the proposed model all spins
are coplanar and the relative angle between nearest-neighbor spins
is $120^o$. In other words, each of the three colors is assigned
to one of three coplanar spin directions (fig.\ref{f.1}). It is
important to note that $ab-initio$ calculations of the molecule in
question have not been performed yet and we have no experimental
evidence on how the plane coplanar to the spins is aligned - it
can be parallel either to one of the pentagons or to one of the
triangles. However, our calculations show that the result is
approximately the same in the both cases.
The octopole moment components in the spherical system can be
found from (\ref{e.2c}) by substituting $n = 3$; $m$ changes from
-3 to 3 by 1, thus yielding seven complex components of the
moment. Our calculations show that the octopole moment components
both in Cartesian and spherical systems are finite but we do not
present here the particular values for the sake of brevity.

Magnetic field with octopolar symmetry at least is required to
measure the octopole magnetic moment components. Such a field can
be generated by an elementary magnetic octopole presented on
fig.\ref{f.2}. The field will create a torque on the Keplerate
molecule(s) placed in the origin of the coordinate system with the
value determined by corresponding octopole moment components. The
estimation of particular torques is beyond the scope of this
article. This approach can be used to study a wide number of
mesoscopic magnets. Neutron spectroscopy can be used to measure
the octopole moment as well and further work in this direction is
required.

\section{"Deoctopolization" in external magnetic field}
Magnetic properties of the \chem{Fe_{30}} molecule and its
multipole moments are discussed in the following. However, the
dependence of the magnetic dipole moment of the magnet in question
on external magnetic field has been quantitatively described in
\cite{b.7} in the framework of the classical statistical physics
(namely, the magnetic moment rises linearly with magnetic field
till the saturation at the critical field value $B_C = 17.7  T$,
what is in perfect agreement with experimental results). The
consideration was based on a study of a three spin system with
antiferromagnetic interactions between the closest neighbour spins
placed in external magnetic field in accordance with the following
Hamiltonian \cite{b.17}:
\begin{equation}
\label{e.4} H = \vec{S_1}\vec{S_2} + \vec{S_2}\vec{S_3} +
\vec{S_3}\vec{S_1} - \vec{\beta}(\vec{S_1} + \vec{S_2} +
\vec{S_3}),
\end{equation}
where $\vec{\beta} = g\mu\vec{B}$/$2JS$ is a dimensionless
magnetic field, $J  > 0$ stands for antiferromagnetic exchange
integral and all spins are of the same length $S$. The Hamiltonian
represents a classical model, i.e. the projections of the spins
are not quantized and can take all intermediate values. The exact
solution for the ground state energy has been found in \cite{b.17}
and it was shown that the ground state energy depended on the
total spin only (or magnetic field, as they are connected
linearly), namely $E_{GS}(\beta) = - 3 / 2 - \beta^2 / 2$. At the
magnetic field value corresponding to $\beta = 3$ all the three
spins are aligned with the magnetic field. The dimension of the
Hilbert space for \chem{Fe_{30}} molecule in the framework of the
quantum model is $6^{30}$ which is far beyond calculation
possibilities of any modern computer. However, an approximate
quantum model based on the isotropic nearest-neighbour
antiferromagnetic Heisenberg exchange is presented in
\cite{b.BandModel} and finally results in appearance of special
small quantum steppes on magnetization typical for molecular
magnets. Analogous consideration would lead to the steppes of the
same kind on the octopole moment components as well but they would
not change the general picture and would be an unnecessary
complication. This is the reason we use classical model here.

\begin{figure}
\twoimages[scale=0.42]{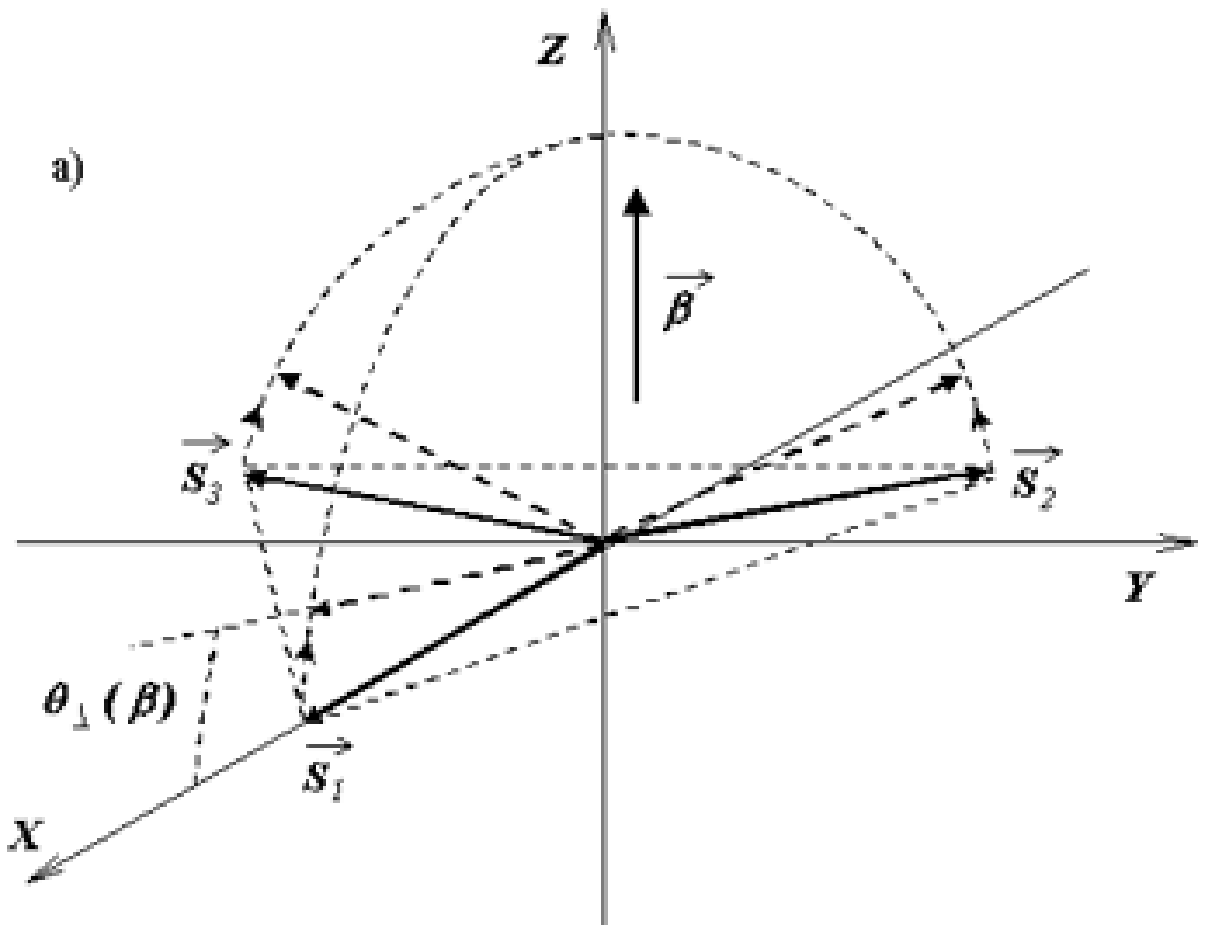}{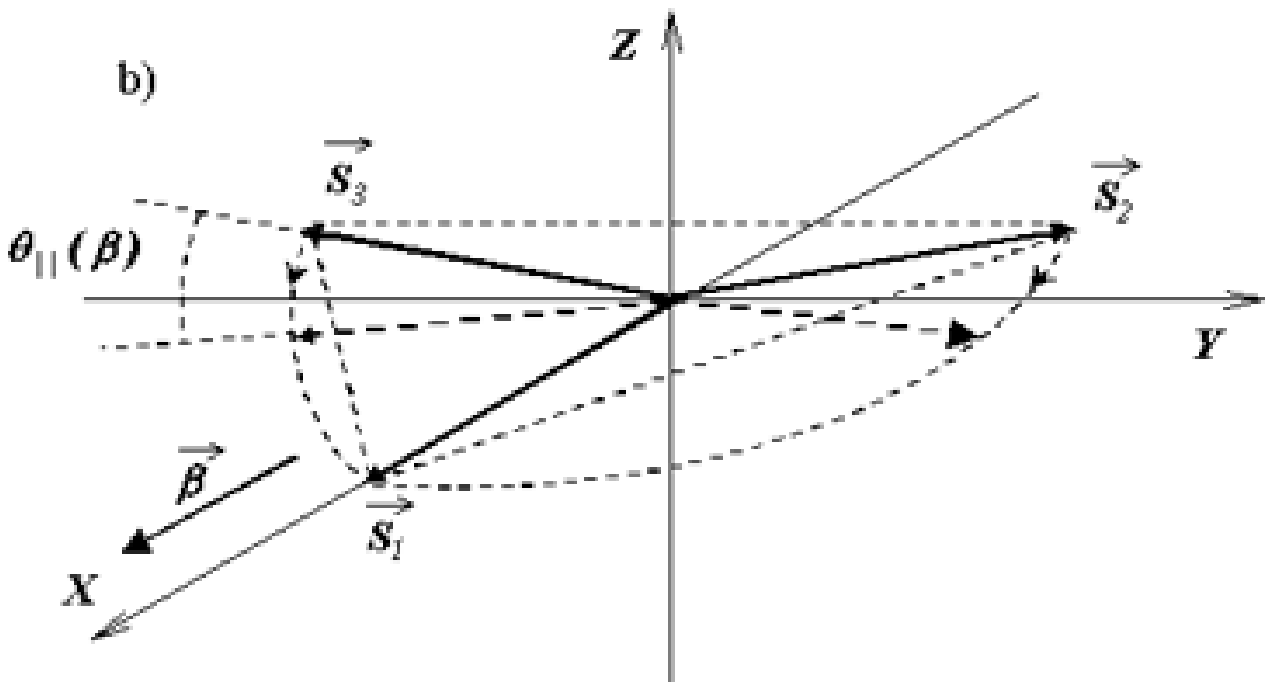} \caption{Two
considered geometries of the \chem{Fe_{30}} molecule magnetization
process: (a) - $\vec{B}$ is perpendicular to the spin plane (case
A); (b) - $\vec{B}$ is parallel to one of the three spins (cases
B, C).} \label{f.3}
\end{figure}

\begin{figure}
\onefigure[scale=0.42]{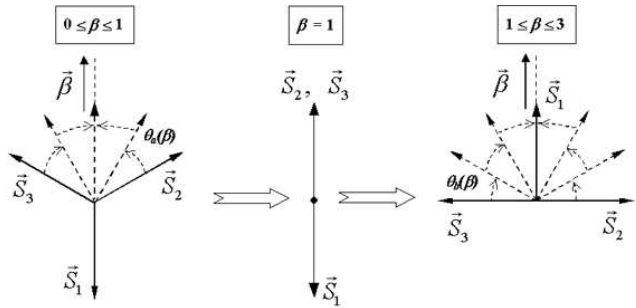} \caption{A simple scheme of the
magnetization process involving the $\vec{S_1}$ spin jump (case
C).} \label{f.4}
\end{figure}

However, due to the antiferromagnetic interactions the three spin
system is highly frustrated and the magnetization process can be
carried out in many ways. The three easiest of them are: A) the
three spins simultaneously leave the plane where they lied before
in the case of magnetic field directed perpendicularly to the spin
plane (fig.\ref{f.3}, a); B) the field is directed along one of
the spins, then the remaining two spins turn in the field's
direction uniformly (fig.\ref{f.3}, b); C) the field is opposite
to one of the spins ($\vec{S_1}$), then the magnetization process
involves two phases separated by the $\vec{S_1}$ spin jump at
magnetic field $\beta = 1$ (fig.\ref{f.4}).

There is one parameter describing each of the first two
magnetization processes: the angle of each spin's declination from
the initial state $\vartheta_\bot(\beta)$ in case (A); the angle
of the $\vec{S_2}$ and $\vec{S_3}$ spins' declination
$\vartheta_{||}(\beta)$ in case (B) and two parameters in the
third case (C) - $\vartheta_a(\beta)$ and $\vartheta_b(\beta)$
(fig.\ref{f.4}). The parameters can be found from minimization of
the system's energy in external magnetic field. This can be done
by projecting the Hamiltonian (\ref{e.4}) on subspaces containing
one or two variables in the first two (A) and (B) and the last
case (C), respectively. Due to simplicity of the considered cases
below we give the answers only:
\begin{equation}
\label{e.5}
\vartheta_\bot(\beta) = \frac{\pi}{2}\beta, \qquad
\vartheta_{||,b}(\beta) = \arcsin\frac{\beta-1}{2}+\frac{\pi}{6},
\qquad \vartheta_a(\beta) =\arcsin\frac{\beta+1}{2}-\frac{\pi}{6},
\end{equation}
where $\beta$ changes from 0 up to its saturation (critical) value
3. Note that the second formula describes two parameters -
$\vartheta_{||}(\beta)$ and $\vartheta_b(\beta)$.

In order to calculate the dependence of high multipole magnetic
moments on the magnetic field we have made an assumption that the
\chem{Fe_{30}} molecule can be divided into 20 triangles of
antiferromagnetically interacting spins (\ref{e.4}) \cite{b.7}.
This assumption can be partially justified by the fact that the
icosidodecahedron is three-colorable without any visible
anisotropy and that this model's results are in perfect agreement
with experiments. Besides, according to recent calculations with
the Density Matrix Renormalization Group technique (DMRG) reported
in \cite{b.18}, the low lying energy levels of the \chem{Fe_{30}}
molecule form a rotational band, i.e. they depend approximately
quadratically on the total spin quantum number $S$. Actually
$\vec{\beta}$ in (\ref{e.4}) should be multiplied by a factor of
$1/2$ when writing the Hamiltonian for all 20 triangles to take
into account that each spin is shared by two neighbouring
triangles. Hence, having the magnetization parameters (\ref{e.5})
we obtain the spin distribution over the molecule at finite
magnetic field and can calculate its multipole moments for any
field value. As we have already noted, the magnetic dipole moment
rises linearly with the field. Surprisingly, the quadrupole
magnetic moment remains equal to zero in all cases (A, B, C) at
any fields up to saturation, and further.

\begin{figure}
\twoimages[scale=0.42]{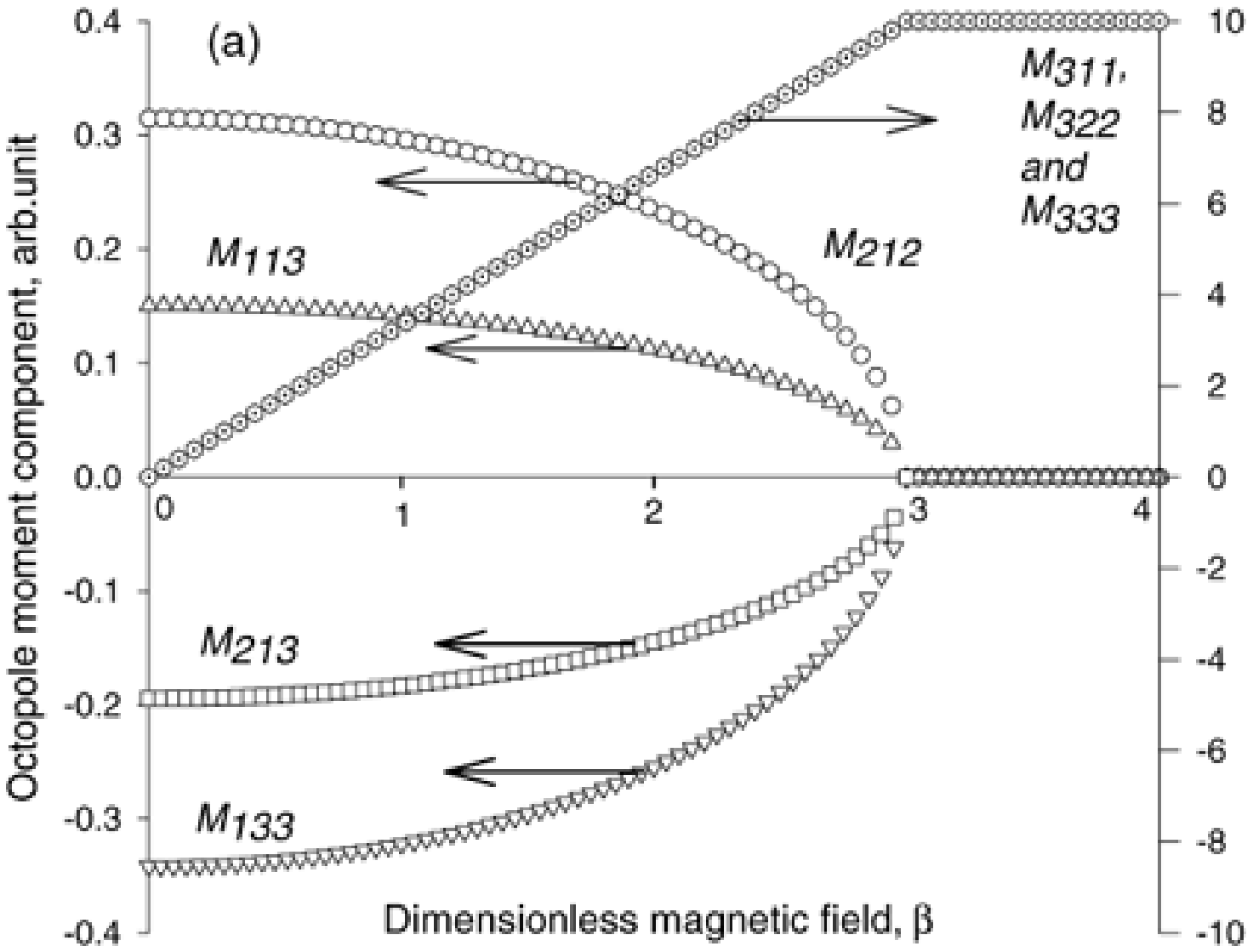}{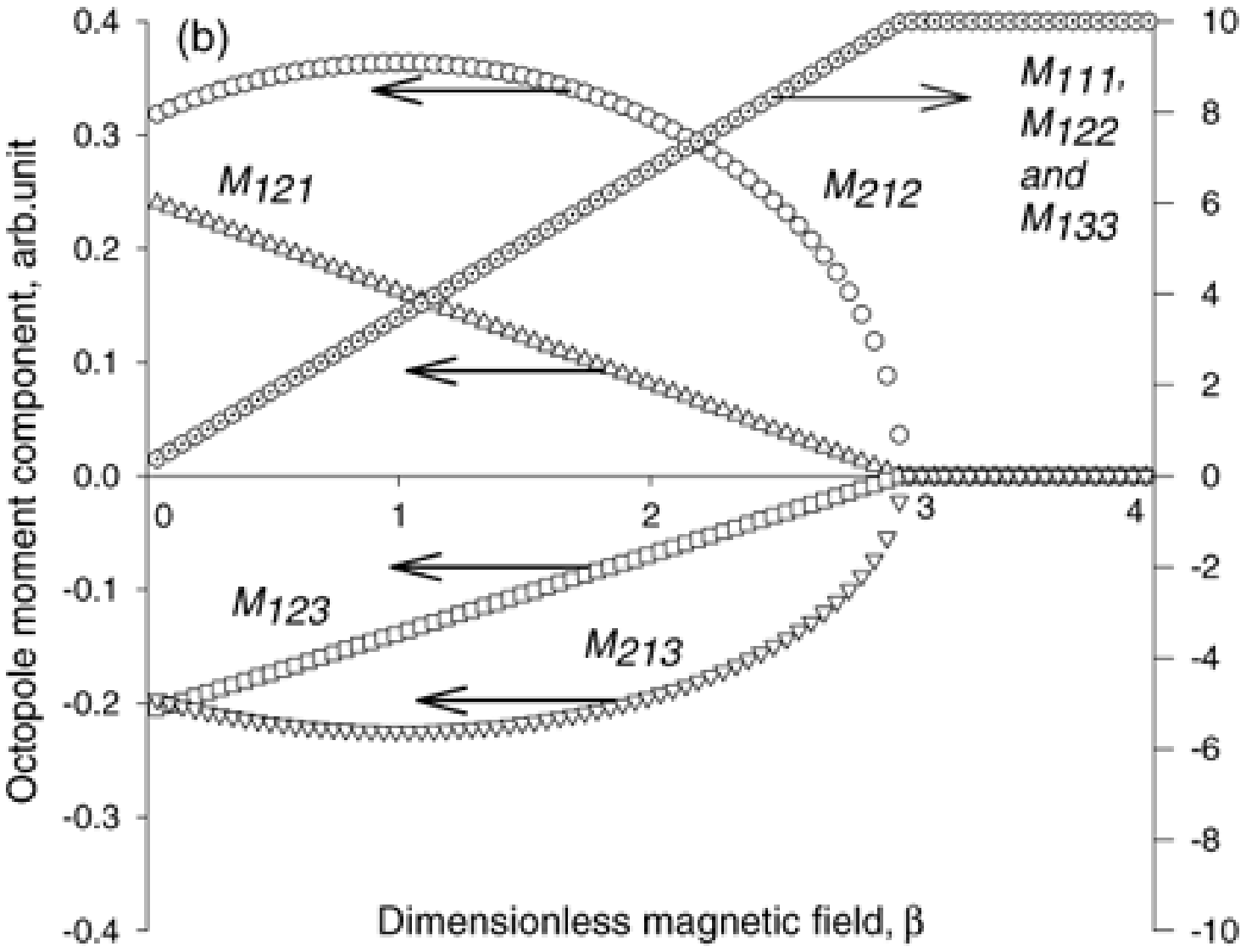} \caption{Basic types
of magnetic field dependence of the components of (a) the octopole
moment components, out-of-plane magnetization (case A) and (b)
in-plane magnetization without the transition (case B)}
\label{f.5}
\end{figure}

The two basic types of the octopole moment components' behavior in
the out-of-plane case (A) are plotted on fig.\ref{f.5}, a. As you
can see, almost all the components vanish with the field. There
are only three linearly raising components. It can be understood
by a simple mathematical consideration. Indeed, at the saturation
all the spins are aligned parallel to the magnetic field (and $Oz$
axis, see fig.\ref{f.3}, a), $i.e$ their $x-$ and $y-$ components
are equal to zero as well as 18 octopole magnetic moment
components $(M_3)_{1ij}, (M_3)_{2ij}, i = 1..3, j = 1..3$
(\ref{e.2b}). Moreover, all components of the kind $(M_3)_{3ij}$
with $i \neq j$ are equal to zero as the contributions from
different spins cancel each other out when we perform the
summation in (\ref{e.2b}). The only three surviving components are
$(M_3)_{311}, (M_3)_{322}$ and $(M_3)_{333}$ as they are the sums
of $squared$ coordinates of all spins multiplied by the $z$
component of the saturated spin and it is obvious that these three
components raise with the field.

\begin{figure}
\twoimages[scale=0.42]{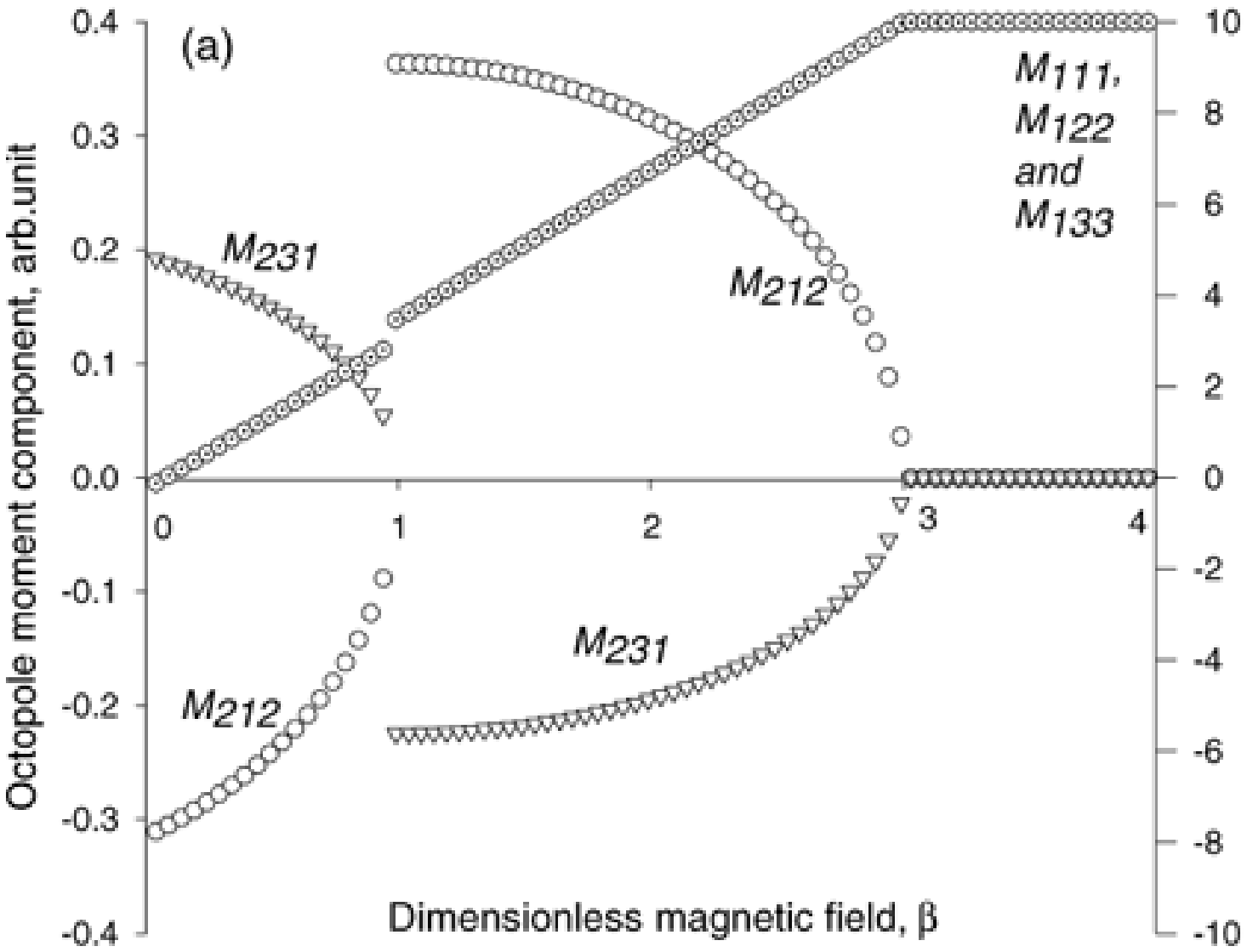}{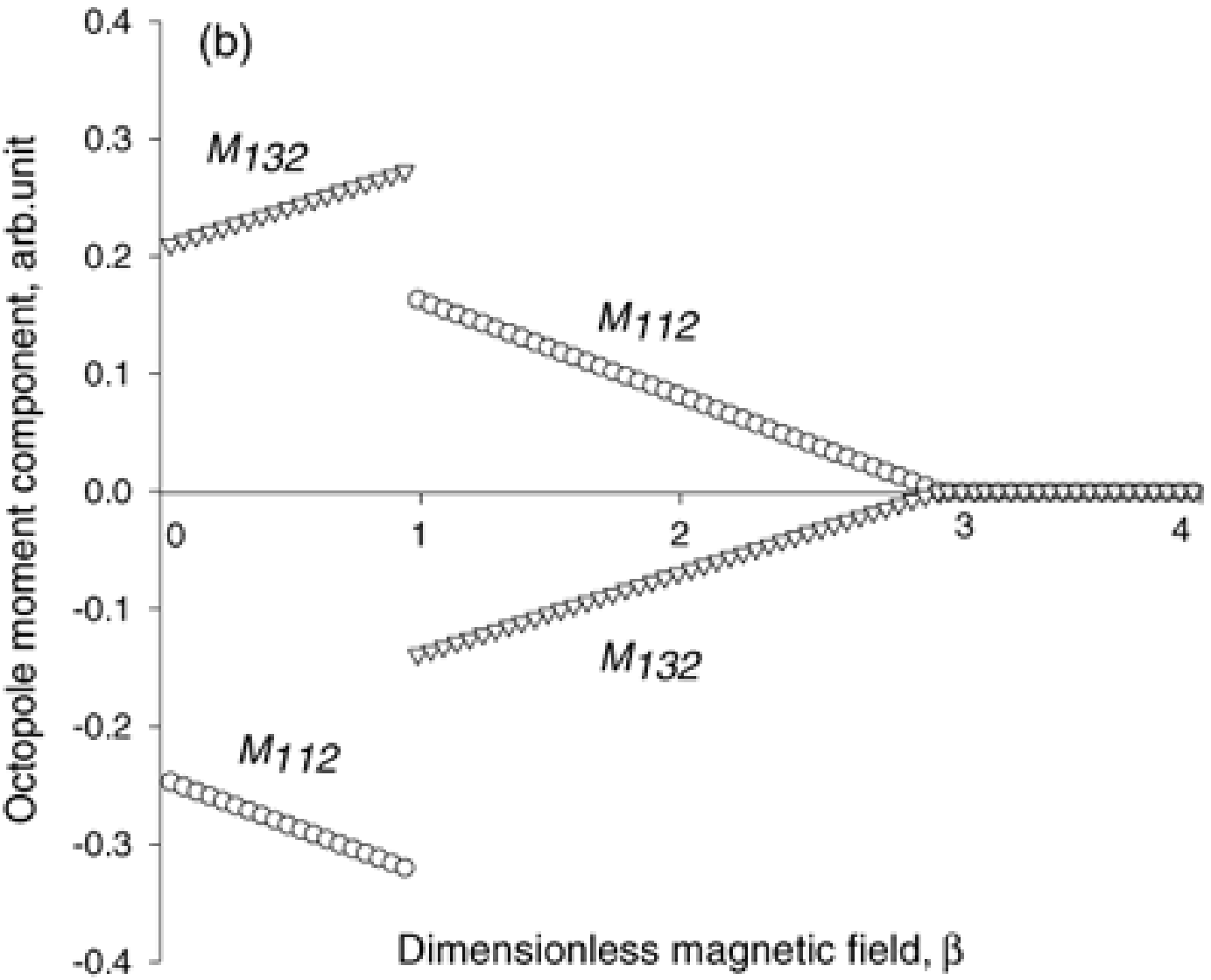} \caption{Four basic
types of magnetic field dependence of the components of the
octopole moment components, in-plane magnetization with the
transition (case C).} \label{f.6}
\end{figure}

However, the in-plane magnetization cases (B, C) are more
interesting. Their general geometry is shown on fig.\ref{f.3}, b.
Note that the magnetic field can have any sign so we cover both B
and C cases with this picture. Nevertheless, the octopole moment
behaves in different ways and we will consider these cases
separately. Take a look at the fig.\ref{f.5}, b. It differs from
fig.\ref{f.5}, a with the curvature of $M_{212}, M_{213}$
components and appearance of dependencies of a new type $M_{121},
M_{123}$. These facts are connected with the way the spin
projections change during the magnetization process (see
(\ref{e.2b}) and fig.\ref{f.3}, b). And just like in the (A) case
we have three linearly growing components - $M_{111}, M_{122},
M_{133}$. The third case (C) is the most interesting as one of the
spins, $\vec{S_1}$ for clarity, abruptly changes its direction for
the opposite one at the magnetic field $\beta = 1$ (see
fig.\ref{f.4}). At the same time, the total spin (and
magnetization) do not experience a jump or even a kink, what
follows from the general solution for the considered problem given
in \cite{b.17}. The octopole moment's components $vs.$ magnetic
field are shown on figs.\ref{f.6} a, b. Quite obviously, all the
dependencies have jumps at the field $\beta = 1$ showing the
abrupt changes of symmetry at this point.

It is important to point out that the three considered cases do
not exhaust all possible ways the magnetization can be realized.
They rather reveal basic features of the "deoctopolization"
process and can be a ground for its future classification.
\section{Conclusion}
The results of the theoretical investigation of the \chem{Fe_{30}}
molecule let us make some important conclusions. First of all it
is shown that dipole, toroid and quadrupole moments of the
\chem{Fe_{30}} molecule are equal to zero. Probably the considered
system has the most symmetric distribution of spin density of all
the mesoscopic magnets synthesized so far. The magnetic octopole
moment is calculated in both Cartesian and spherical coordinates.
A model of an elementary magnetic octopole as a possible
experimental unit for octopole magnetic moment component
measurement is proposed. Toroid and quadrupole moments are shown
to remain equal to zero at any magnetic fields while components of
the octopole moment are changed substantially. Different
geometries of the magnetization process are considered and
discussed.
\acknowledgments This work is supported by INTAS (grant
99-01-839) and RFBR (grants 02-02-17389 and 01-02-17703).

\end{document}